\documentclass[namedreferences]{solarphysics}
\usepackage[optionalrh]{spr-sola-addons}
\usepackage{epsfig}
\usepackage{graphicx}
\usepackage{color}
\usepackage{url}
\newcommand{\etal}{{\it et al.}}
\usepackage[pdfborder={0 0 0 },urlcolor=blue]{hyperref}
\ifx \doiurl \undefined \def \doiurl#1{\href{http://dx.doi.org/#1}{\url{#1}}}\fi
\ifx \adsurl \undefined \def \adsurl#1{\href{http://adsabs.harvard.edu/abs/#1}{\url{#1}}}\fi

\begin{document}
\begin{article}
\begin{opening}

\title{Short-Term Variations in the  Equatorial  Rotation 
Rate of Sunspot Groups} 

\author{J.\ Javaraiah$^1$ and L.\ Bertello$^2$}  
\institute{$^1$\#58, 5th Cross, Bikasipura (BDA), Bengaluru-560 111,  India.\\
Formerly with Indian Institute of Astrophysics, Bengaluru-560 034, India.\\
email: \url{jajj55@yahoo.co.in;  jdotjavaraiah@gmail.com}\\
{$^2$National Solar Observatory, Tucson, U.S.A.}\\
email: \url{lbertello@nso.edu}\\
}

\runningauthor{J.\ Javaraiah and L.\ Bertello}
\runningtitle{Variation in the Sun's Equatorial Rotation Rate}

\begin{abstract}
We have detected several periodicities in the solar equatorial rotation rate
  of sunspot groups in
the Greenwich Photoheliographic Results (GPR) during the period 1931\,--\,1976,
the {\it Solar Optical Observing Network} (SOON) during the period 
1977\,--\,2014, 
and the Debrecen Photoheliographic Data (DPD)  during the period 1974\,--\,2014.
We have compared the results from the fast Fourier transform (FFT),
 the maximum entropy method (MEM), 
 and from Morlet wavelet power-spectra of the equatorial rotation 
rates  determined from  SOON  and DPD sunspot-group data during the
 period 1986\,--\,2007   with those of the Mount Wilson  Doppler-velocity 
data during the same period  determined  by
 Javaraiah \etal (2009, {\bf 257}, 61).
We have also compared the power-spectra computed from 
the DPD and the combined  GPR and SOON sunspot-group data during the
 period 1974\,--\,2014 to those from the GPR sunspot-group data during
 the  period 1931\,--\,1973. Our results suggest  a
 $\sim$250-day period in the equatorial
rotation rate determined from both the
Mt. Wilson Doppler-velocity data and the sunspot-group data 
during 1986\,--\,2007.
 However, a wavelet analysis reveals that this periodicity 
  appears mostly  around  1991  in the velocity data, while
 it is present in most of the solar cycles covered by the sunspot-group  
data,  mainly near the  minimum  epochs of the solar cycles.
We also found the signature of a period of $\sim$1.4 years
 period in the velocity data
during 1990\,--\,1995, and in  the equatorial rotation rate of sunspot
 groups mostly around the year 1956.  
The equatorial rotation rate of sunspot groups reveals 
a  strong $\sim$1.6-year periodicity around
1933 and 1955  and a weaker one
around  1976, and a strong $\sim$1.8-year periodicity 
around 1943.  Our analysis also suggests 
periodicities of $\sim$5 years, $\sim$7 years, and  $\sim$17 years
as well as 
some other short-term periodicities. However, short-term periodicities 
are  mostly present at the time of solar minima. Hence,  short-term 
 periodicities  cannot be confirmed because of 
 the larger uncertainty in the data.  

\end{abstract}
\end{opening}

\section{Introduction}
 In addition to the well-known 11-year  solar cycle, 
solar activity varies on many shorter and longer timescales. For example,
 periodicities shorter than a year and of about two years 
  are found in many 
solar activity  indices (\opencite{rieg84}; \opencite{lean89}; \opencite{pap90};
 \opencite{bai91}; \opencite{bou92}; \opencite{rich94}; \opencite{kriv02};
 \opencite{ozgu03}; \opencite{kane03}; \opencite{bai03}; \opencite{obri00};
 \opencite{obri07}; \opencite{chow09}; \opencite{scaf13}; \opencite{kilc14}; 
\opencite{chow16}; and  references therein). Studies of similar variations 
in the solar rotation data  may help us to better understand the physical
 processes
responsible for the solar variability.
Recently, \inlinecite{jj13} determined solar cycle variations in the 
mean equatorial rotation 
rate of the sunspot groups, with and without the data 
of  abnormal angular
 motions  of  the sunspot groups. 
He found  a large difference between the solar cycle variations in the 
yearly mean values of the equatorial rotation rates
 determined from the Mt. Wilson Doppler-velocity data and the   
sunspot-group data that did not include   the abnormal motions.
 The patterns of the solar cycle variations of the 
 equatorial rotation rate determined from the sunspot-group data that 
included  the abnormal angular motions of the sunspot groups and  
 the Mt. Wilson Doppler-velocity data closely resemble
 each other.  Earlier, \inlinecite{jk99} analyzed the Mt. Wilson 
Doppler-velocity 
data during 1986\,--\,1994 and found a $\sim1.2$-year and a few
 other short-term periodicities in the  mean solar rotation rate.  
A study of the same data by \inlinecite{jj11} found a
 few quasi-periodicities, ranging from 
a few days to a month, in
the solar differential rotation rate. These short-term periodicities in
the solar surface equatorial rotation rate were  also found
 by \inlinecite{jub09}, using
the corrected Mt. Wilson Doppler-velocity measurements during the period 1986\,--\,2007.
Here we investigate the possible  short-term  periodicities in 
the mean equatorial rotation rate of sunspot-group data and compare  
these periodicities to those found in the equatorial rotation rate 
determined from the Doppler-velocity data. Because of
some inconsistencies in the rotational results obtained from the sunspot data
measured at different observatories~(\opencite{jbu05}, and references therein),
here we use three sets of sunspot-group data.
The Mt. Wilson Doppler-velocity data were acquired from  December 1986 to March 2007, during Solar Cycles 22 and 23, while the
sunspot data are available for a much longer period of time. 

The data analysis is described in the next section, while in Section~3 we
 discuss the results from the spectral and wavelet analysis of the
Mt. Wilson Doppler-velocity data and the three sets of sunspot-group data. 
A comparison
of power spectra with those previously determined from the Mt. Wilson 
Doppler-velocity data by \inlinecite{jub09}
is also shown. The summary and discussion of these results is given in
 Section~4.

\section{Data Analysis}
The solar differential rotation can be determined
from the full-disk velocity data  using the standard polynomial
expansion 
$$\omega(\phi) = A + B \sin^2 \phi + C \sin^4 \phi ,\eqno(1) $$
 while for sunspot data,  which are confined to only low and medium latitudes,  
 it is sufficient to use the first two terms of the expansion, $i.e.$
$$\omega(\phi) = A + B \sin^2 \phi ,\eqno(2) $$

\noindent where $\omega(\phi)$ is the solar sidereal angular velocity at
latitude $\phi$, the coefficient  $A$  represents
the equatorial rotation rate and $B$ and $C$  measure the
latitudinal gradient in the rotation rate, 
 $B$ is associated mainly with low latitudes and $C$ is associated largely 
with higher latitudes.

Here we use the Greenwich Photoheliographic Results, GPR (1931\,--\,1976), SOON (1977\,--\,2014),
 and Debrecen Photoheliographic Data, DPD (1974\,--\,2014), sunspot-group data.
 The GPR and SOON data were taken
from    the website  
{\tt http://solarcience.msfc.nasa.gov/greenwich.shtml}, and the DPD data
 from {\tt http://fenyi.solarobs.unideb.hu/pub/DPD/}. For each daily 
sunspot-group observation, the
files include information about 
the time of the observation, the heliographic latitude ($\phi$) 
and longitude ($L$), and the central meridian distance (CMD).  

The GPR data  
have been compiled from the majority of the 
white-light photographs stored at the 
Royal Greenwich Observatory  and at the Royal Observatory at the Cape of Good 
 Hope. The gaps in these 
observations were filled with photographs from other observatories, such as 
Kodaikanal Observatory, India, the Hale Observatory, California, and 
the Heliophysical Observatory at Debrecen, 
Hungary. The Royal Greenwich Observatory terminated the 
 publication of GPR 
at the end of 1976. Since 1977 the Debrecen Heliophysical Observatory 
took over this task.
The SOON data  included measurements made   by the
United States Air Force (USAF) from
the sunspot drawings of a network of  observatories
 that  included telescopes
in Boulder, Colorado, Hawaii, $etc$.
 David Hathaway  scrutinized
the GPR and SOON sunspot-group  data and produced a reliable
continuous data series
from 1874 until today (\opencite{hath03}; \opencite{hath08}; \opencite{hath15}).
 The DPD contain the positions and areas of sunspots, the total area 
and the mean positions of the sunspot groups, for each day 
compiled by using white-light full-disk observations taken at the 
Heliophysical Observatory, 
Debrecen, Hungary, and its Gyula Observing Station as well as at some 
other observatories. When no ground-based observation was found,
 space-borne quasi-continuum images obtained by the {\it Michelson Doppler
 Imager} (MDI) onboard  the Solar and Heliospheric Observatory (SOHO)
 were used~(for details see \opencite{gyr10}). 

The determination of the sidereal rotation rates of the sunspot groups is briefly described in 
 \inlinecite{jj13}. The ratio of the difference ($\Delta{L}$) between
  the values of heliographic longitudes  to
the difference  ($\Delta t$) 
 between  the times of the consecutive day observations of 
the sunspot groups is first computed. Then,  the value of the Carrington 
rigid-body rotation rate 
($14.18^\circ$ day$^{-1}$) is added to this ratio. Javaraiah and co-workers have used
 this method in most of their earlier
 studies of solar rotation  rate determined from sunspot-group data 
(\opencite{jbu05}; \opencite{jj13}, and references therein).
 The data (the values  of $\omega$  and  mean $\phi$  of the  
  consecutive days)  are fitted to Equation~(2)  to  obtain the values of 
$A$  and $B$ coefficients.  
 In this article we mainly intend to compare 
the power spectra of the equatorial rotation rate [$A$] derived from the
 sunspot-group data with the corresponding spectra derived from 
 the Mt. Wilson Doppler-velocity data by \inlinecite{jub09}.
 The daily values of
 $A$ determined from the corrected (for scattered light, $etc.$,
 \opencite{ul01}) Mt. Wilson Doppler-velocity measurements cover the 
time interval 
from 1986 to 2007. In an earlier analysis (\opencite{jub09}) of this data set,  
large spikes, $i.e.$ the values $>2\sigma$ (where $\sigma$ is the standard 
deviation), were removed from the daily data. 

The Mt. Wilson Doppler-velocity daily data set contains several missing days,
with gaps as long as 
49 days.  \inlinecite{jub09} binned these data  in 61-day time intervals
 to produce
a time series without gaps that is suitable for the required spectral analysis. 
Here we use a similar
approach for the sunspot-group data by binning them into 61-day time intervals.
First we analyzed  the SOON and DPD sunspot-group data during 1986\,--\,2007, 
 $i.e.$, for the same period for which the Mt. Wilson velocity-data
 are available. The coalignment in time of the sunspot-group data with the
velocity data produced 126  61-day samples.  
 Both the DPD sunspot-group data and the combined GPR and SOON (GPR-SOON) data
 are available for the 
period 1974\,--\,2014. We compared the power spectra of the $A$ times series
determined from these two data sets and also   
analyzed the  GPR  sunspot-group data from
{\tt http://fenyi.solarobs.unideb.hu/pub/DPD/} for the period  1931\,--\,1973.
GPR sunspot-group data before 1931 were not used because 
of their poor quality during periods of solar cycle minima.

We did not use the sunspot-group data with
$|CMD|>75^\circ$  
on any day of the sunspot-group life time. This reduces the foreshortening 
effect (if any). However, in this analysis the abnormal motions  of the  
sunspot groups, $i.e.$ the data      
 corresponding to $\delta L > 3^\circ$ day$^{-1}$,  were included. 
This increases the
amount  of data  in a given 61-day interval, 
particularly during the solar cycle minima. In addition, these data
were found to agree quite well with the Doppler-velocity data (\opencite{jj13}).
Equation~(2) was used to derive the values of $A$ from the sunspot-group data,
 and then the FFT, MEM, and Morlet wavelet power spectra of the $A$ time series 
 were computed.  The MEM  FORTRAN code  was provided
  by  A. V. Raveendran,
which was also used in the earlier papers~\cite{jg95,jg97a,jub09}.

Both FFT and MEM spectral methods are well-known techniques. In particular,
MEM uses a 
parametric modeling approach to estimate the power spectrum of a time 
series. The method is data adaptive because it is based upon an
 autoregressive (AR) modeling process. An AR process is predictive;
 any point (after the first) 
is calculated by a linear combination of $M$ (order of the process) previous values. 
An important step in this method is the optimal selection 
of the order $M$. If $M$ is chosen too small, then the model smooths the data excessively and the resulting estimate of the power spectra is poorly resolved. 
If $M$ is chosen to be too large, then frequency shifting and spontaneous 
splitting of peaks can occur (see \opencite{ub75}, and references therein).       

Before the FFT is computed, the mean value is subtracted from the data, 
and the time series tapered by multiplying the first and the last 10\% of the data 
by a half cosine bell~(\opencite{bw71}). We have padded the time series 
with  zeros so that the number of data points [$N$] corresponds to an 
exact power of two (we have added two zeros to the data during 
1986\,--\,2007).  The significance levels of the peaks in the  FFT spectra
are computed by using  both white-noise and red-noise 
models (see \opencite{torr98}). 
The MEM code that we have used here takes the values for $M$
 in the range [$N/3$, $N/2$] (\opencite{ub75})   or 2N/$\ln$(2N)
(\opencite{berry78}). We adjusted the value of $M$ until
the FFT and MEM spectra showed a good agreement with each other.
We found that $M = N/3$ is suitable for our analysis, producing
spectra with peaks that are considerably sharp
and well separated. 

The wavelet transform is particularly effective for analyzing
 non-stationary signals. It can detect transient periodic signals
and track their amplitude variation in time, as  is typical of
most solar phenomena. Here we applied the Morlet
wavelet analysis (\opencite{torr98}) to the $A$ time series.

\begin{figure}
\centerline{\includegraphics[width=\textwidth]{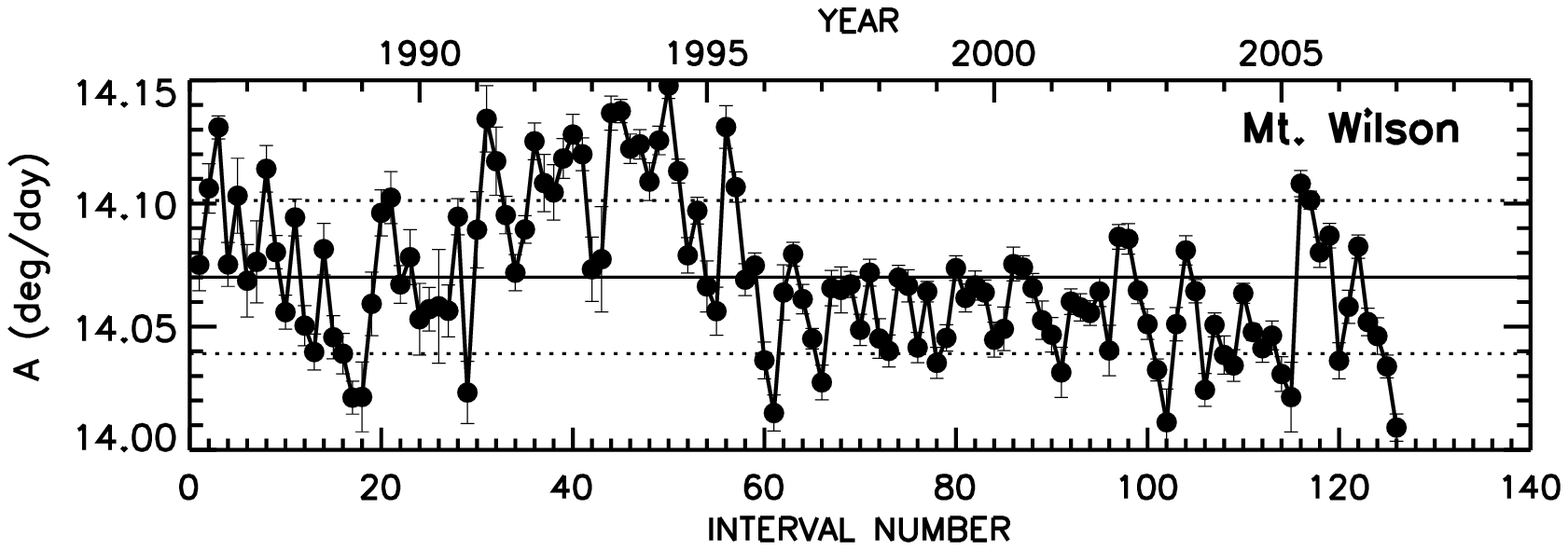}}
\centerline{\includegraphics[width=\textwidth]{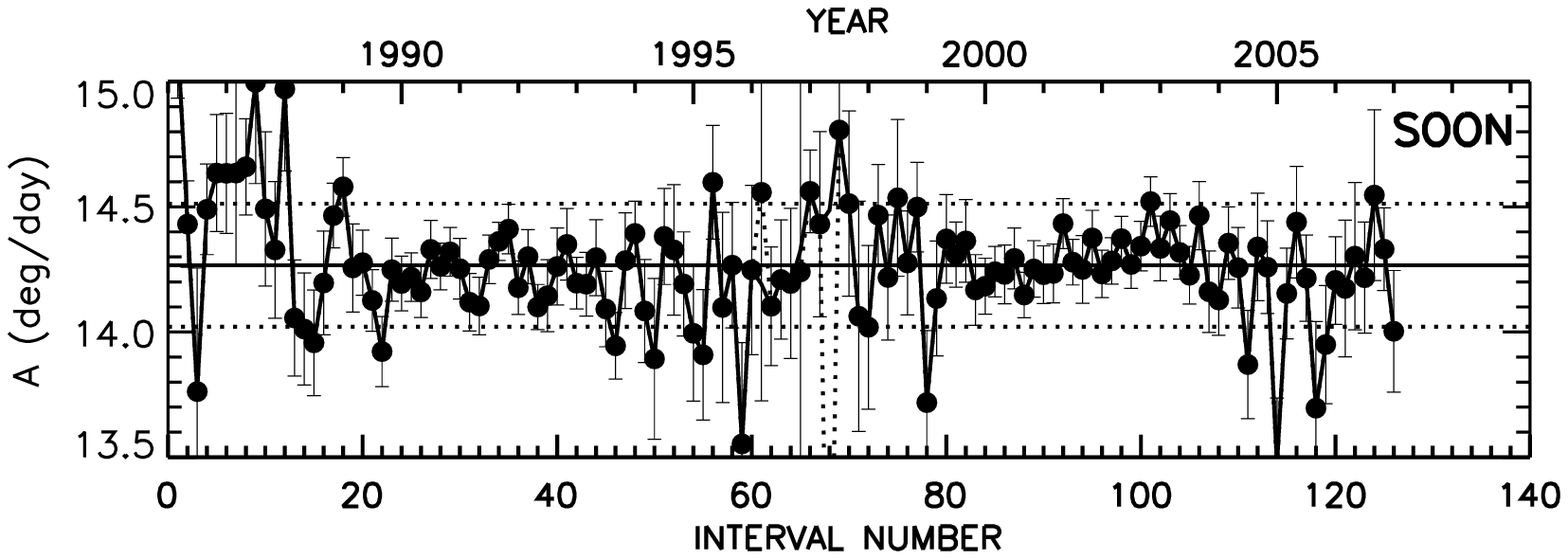}}
\centerline{\includegraphics[width=\textwidth]{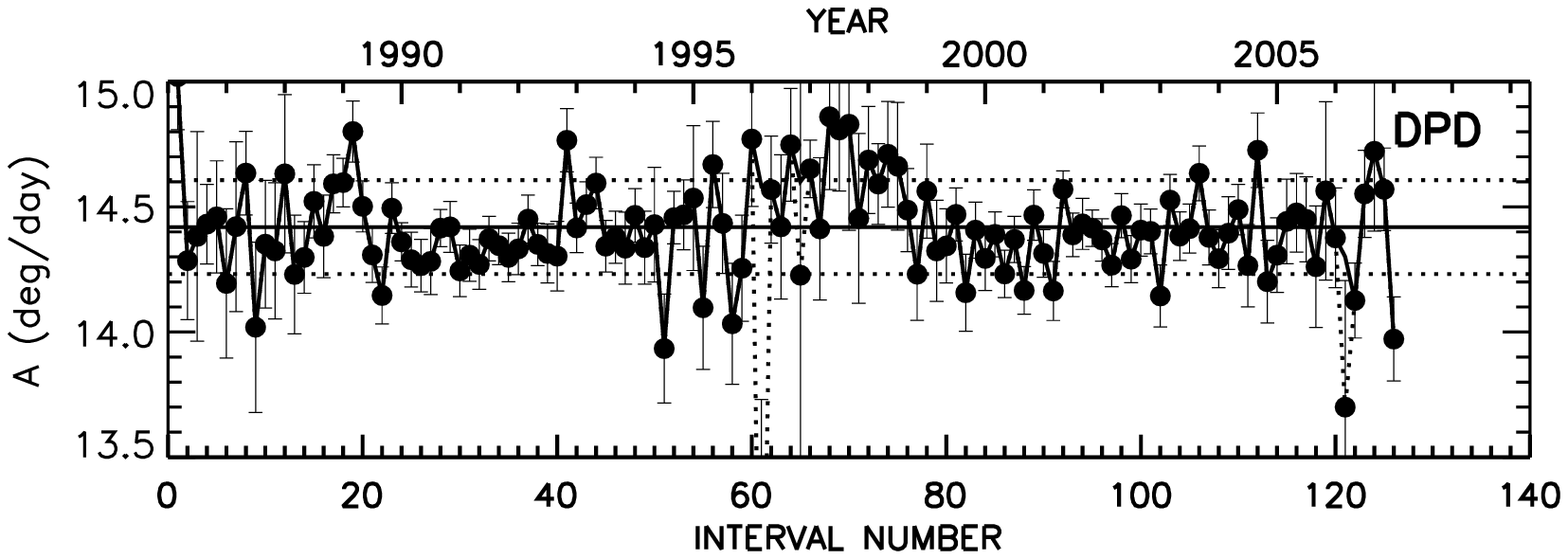}}
\caption{Plots of $A$ in 61-day intervals determined from Mt. Wilson
 Doppler-velocity data (upper panel),  SOON (middle panel), and 
 DPD (lower panel) sunspot-group data  during the  period  
 1986\,--\,2007 $versus$ time (interval numbers). Error-bar 
represents the standard error in case of the velocity data and 
standard deviation ($\sigma$) in case of the sunspot-group data.  
The horizontal continuous line represents the  mean and 
the horizontal dotted lines indicate the corresponding 
root-mean-square deviations.
In case of $A$ determined from the sunspot-group data, the values whose
 $\sigma$ values  exceeded by 2.6 times the corresponding median value are 
replaced with the average of the corresponding values and their respective two
 neighbors.  The continuous curve represents the corrected data,  and 
the original data points are connected by the dotted curve.}
\label{Fig.1}
\end{figure}

\begin{figure}
\centerline{\includegraphics[width=\textwidth]{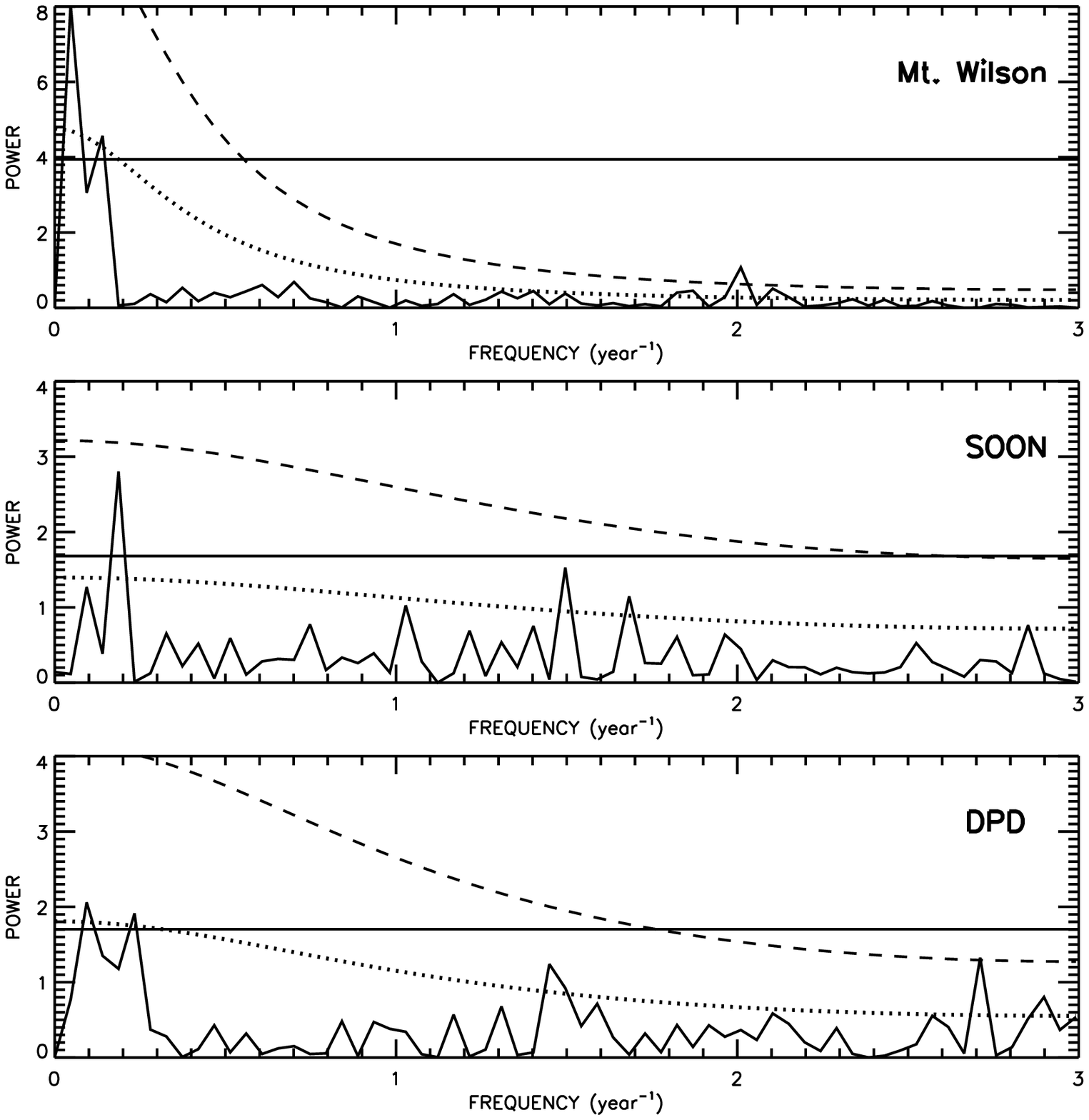}}
\caption{The upper panel shows the FFT  spectrum of $A$ computed for 
the 61-day intervals
 of the corrected  Mt. Wilson Doppler-velocity data after subtracting 
the cosine fit of the one-year period.  The middle and lower panels are 
  the FFT spectra of $A$ determined from the corrected 61-day binned SOON 
and DPD sunspot-group data, respectively. 
 The continuous  horizontal lines are drawn at the $3\sigma$ levels 
of the  power in the respective spectra. The dotted and dashed curves
 represent the mean and 90\% significant  red-noise spectra of $A$  
determined from the Mt. Wilson velocity-data (the assumed 
$\mathrm {lag}$-1 auto-correlation $\alpha =0.653$), 
SOON ($\alpha = 0.165$) and DPD ($\alpha = 0.288$)  sunspot-group data.}
\label{Fig.2}
\end{figure}

\section{Results}

The upper panel of Figure~1 shows the variations
 in $A$  determined 
 from     the Mt. Wilson velocity-data 
 during 1986\,--\,2007. The middle and  lower panels show   
  the corresponding variations determined  
 from SOON and DPD sunspot-group data for the same period.
Both median and standard deviation values were determined for each time series
 and were used
to identify outliers. 
Values of  $A$ whose $\sigma$ values 
 exceed  2.6 times the corresponding median value were 
replaced with average of the corresponding values and their respective
 two neighbors. The solid curve represents the corrected data  and 
the original data points are connected by the dotted curve. 
The three plots in Figure~1 show  variations on different
 timescales, ranging from a few months to several years,
in the $A$ coefficient determined from  
 the velocity and the sunspot-group data.
However, the correlation between the velocity data and the sunspot-group
 data is only 
-6\% to -8\%, while the correlation 
between the variations in $A$ determined from  SOON and DPD sunspot-group data
is much higher, around 33\%.
The large variation in the velocity data during 1990\,--\,1995 may be caused by  
several changes in the Mt. Wilson spectrometer that occurred during this period (see \opencite{jub09}). 
 Figures~2, 3, and 4  show the corresponding
  FFT, MEM, and Morlet wavelet spectra of the data.
In the wavelet spectra, the cross-hatched regions indicate the cone of
influence where the edge effects become significant and the signal
is statistically unreliable~(\opencite{torr98}).
The dashed contours show the power above the 95\% confidence level.
The FFT and MEM spectra of the velocity data shown here are 
 slightly different from those discussed in  \inlinecite{jub09}.
For the reasons discussed in Section~1, the high-frequency peaks that were
detected in \inlinecite{jub09} 
are much more reduced in the present spectra.
  However, the overall properties of the present spectra are still 
 consistent with the conclusions and discussion in that early article.

 Figures~2 and 3 show  a  considerable similarity 
in both the FFT and MEM power spectra  
of the $A$ coefficient determined from the velocity and the sunspot-group data. 
That is, in the vicinity (within the uncertainty limit) 
 of most of the peaks  in the FFT and MEM  spectra 
there are peaks in the
 corresponding spectra of $A$ determined from the  
sunspot-group data. Only a few peaks present in one spectrum are
 absent in the other (this can be seen easily in the MEM spectra.) 
The correlation between the FFT spectra of the velocity data and 
SOON sunspot-group data is poor (3.3\%), but the correlation 
between  the  FFT spectra of velocity data and DPD sunspot-group data is
much higher (35\%), and it is  even better than  the correlation (27\%) between the
 FFT spectra of  SOON and DPD sunspot-group data. 
The values of the correlations between the corresponding MEM spectra 
are found to be 23\%, 25\%, and 20\%,  respectively.  
The 2.1-year and 156-day peaks are present in the spectra of
 the velocity data, but seem to be absent from the spectra of 
 the sunspot-group data. 
The 9\,--\,13 year, 350-day\,--\,1.01-year,
 221\,--\,228 day, $\sim$135-day, and 
$\sim$127-day peaks seem to be present in the 
spectra of the sunspot-group data, but not in the
 spectra of the velocity data. Overall, the spectra of 
 the velocity and the sunspot-group data show periodicities of 5\,--\,7 years, 
$\sim$1.4 years, 241\,--\,248 days, $\sim$199 days,  $\sim$182 days, and 
 142\,--\,155 days 
 in the solar equatorial rotation rate (the 182-day periodicity may be caused 
by the seasonal effect).  However,   except for the 
5\,--\,7-year peaks, none of other peaks are significant 
in the FFT spectra at a 95\% confidence level.

The values of the assumed $\mathrm {lag}$-1 auto-correlation  ($\alpha$) 
of the red-noise models (\opencite{torr98}) of the FFT spectra 
of the velocity  and the 
sunspot group-data are 0.653 and 0.165, respectively 
 (note that  $\alpha = 0$ yields white-noise spectrum).
 In the case of the red-noise model,
 for a peak to be significant at a given significance level, a higher 
value of the power is required than for the white-noise background 
spectrum (\opencite{torr98}). Except for the $\sim$135-day peak in the FFT 
spectrum of DPD sunspot-group-data, none of the other peaks is
significant at a 90\% confidence level in the red-noise model of
 the FFT spectrum of  $A$. Only a peak at a period of  
$\sim$182 day period is significant in the FFT spectrum of the velocity data.
Therefore, no significant periodicities are detected from this analysis.

\begin{figure}
\centerline{\includegraphics[width=\textwidth]{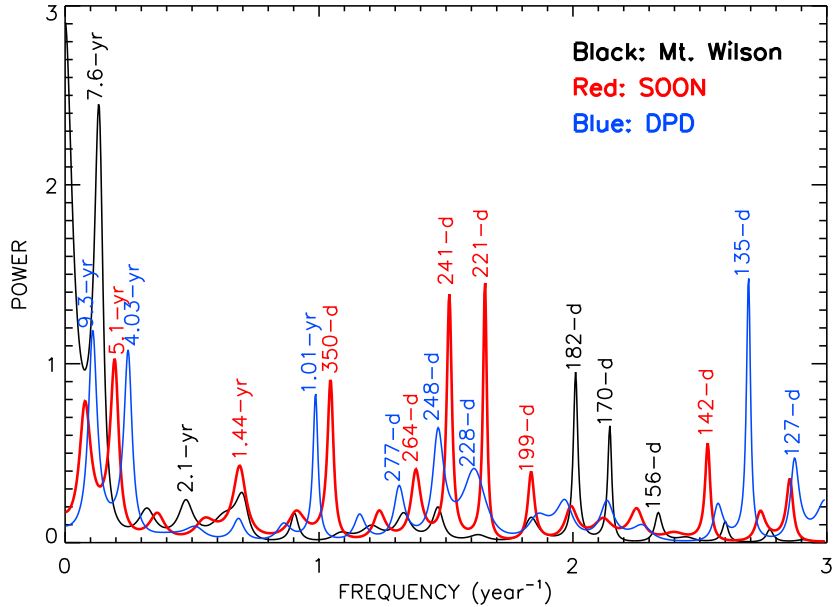}}
\caption{MEM power spectra of $A$ determined from the corrected
  Mt. Wilson Doppler-velocity data, SOON, 
 and DPD  sunspot-group
 data  shown in Figure~1. The value of the determined period is given for each
 well-defined peak.}   
\label{Fig.3}
\end{figure}

Figure~4 seems to suggest  a strong periodicity of $\sim$248 days
 ($\sim$0.69 years)  in 
 the time series of $A$ determined from both the Mt. Wilson Doppler-velocity 
 data and sunspot-group data,  but occurring at
different temporal epochs. That is,  this periodicity
appears in the velocity data around 
1991, while it appears 
 in  the  sunspot-group data around 1995. This periodicity appears 
in both SOON and DPD sunspot-group data 
  around  1986 and 2005 (it appears to be strong in SOON data).
 That is, 
this periodicity seems to  exist at $\sim$11-year intervals 
in the $A$ coefficient of the sunspot-group data,  mainly  near the  
minimum epochs of  solar cycles. Figure~4 also shows 
a $\sim$182-day periodicity around 
  1987, 1995, and 2005 in  both the velocity and 
 the sunspot-group data. 
A weaker $\sim$1.4-year signal is present
in the spectrum of the  velocity data  during 
the period 1989\,--\,1995, while  
 in the spectrum of $A$ determined from the  sunspot-group data 
the same periodicity 
  is present  around 1988 and during  1997\,--\,1999. 
A weak and slightly longer (1.6\,--\,1.7-year) 
periodicity seems to exist after  2002 in the velocity data, while 
 a $\sim$5-year periodicity appears 
 in the $A$ time series of  the sunspot groups  almost  throughout
 the period 1986\,--\,2006.
There is also a weak signal with a 3\,--\,4-year periodicity  in the
sunspot-group data  mainly during Cycle~22. 
The temporal dependence of the $\sim$7.6-year periodicity found in 
the FFT and MEM analyses of the velocity data is not detected here.

\begin{figure}
\centering
{\includegraphics[width=10.5cm]{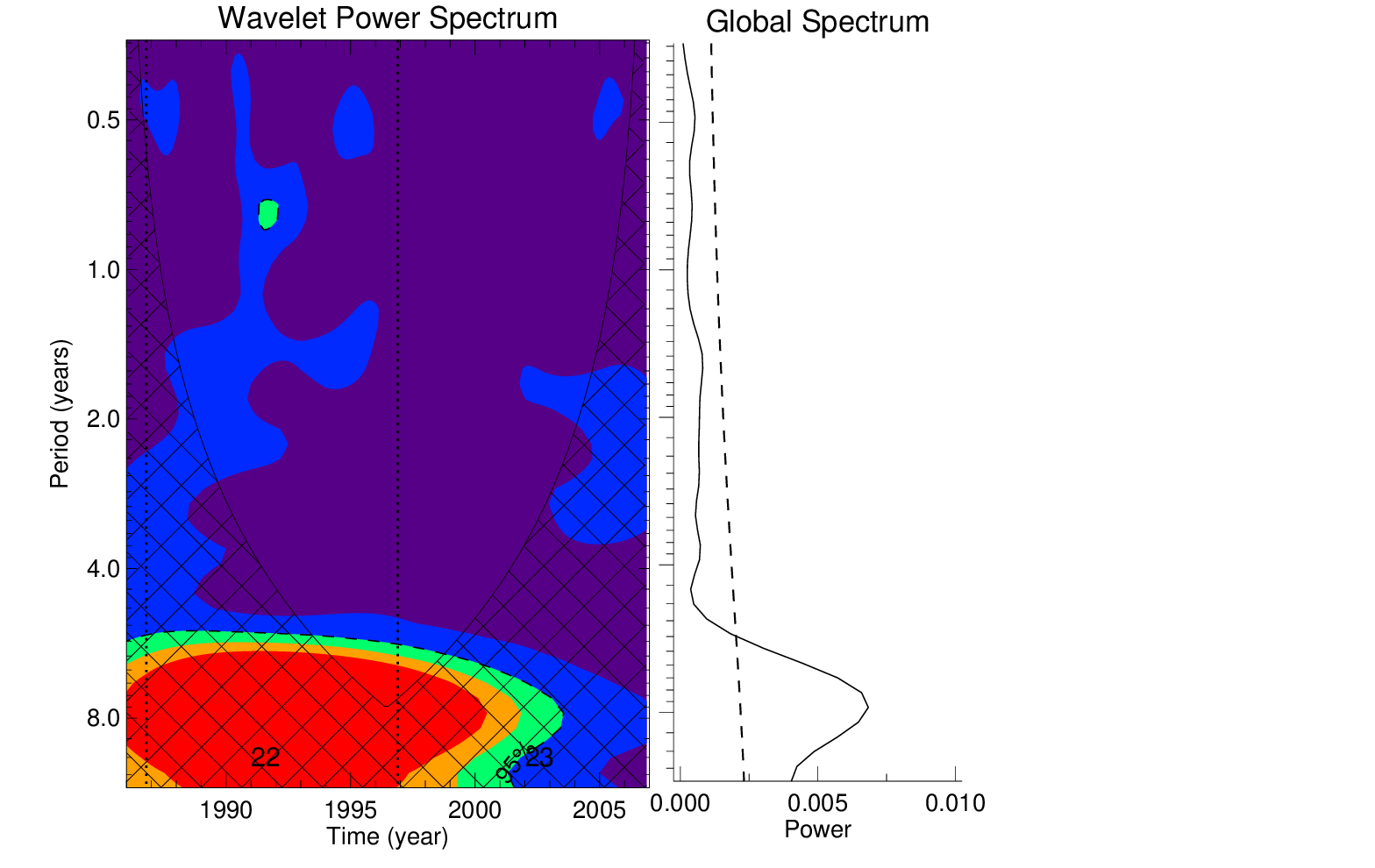}}
{\includegraphics[width=10.5cm]{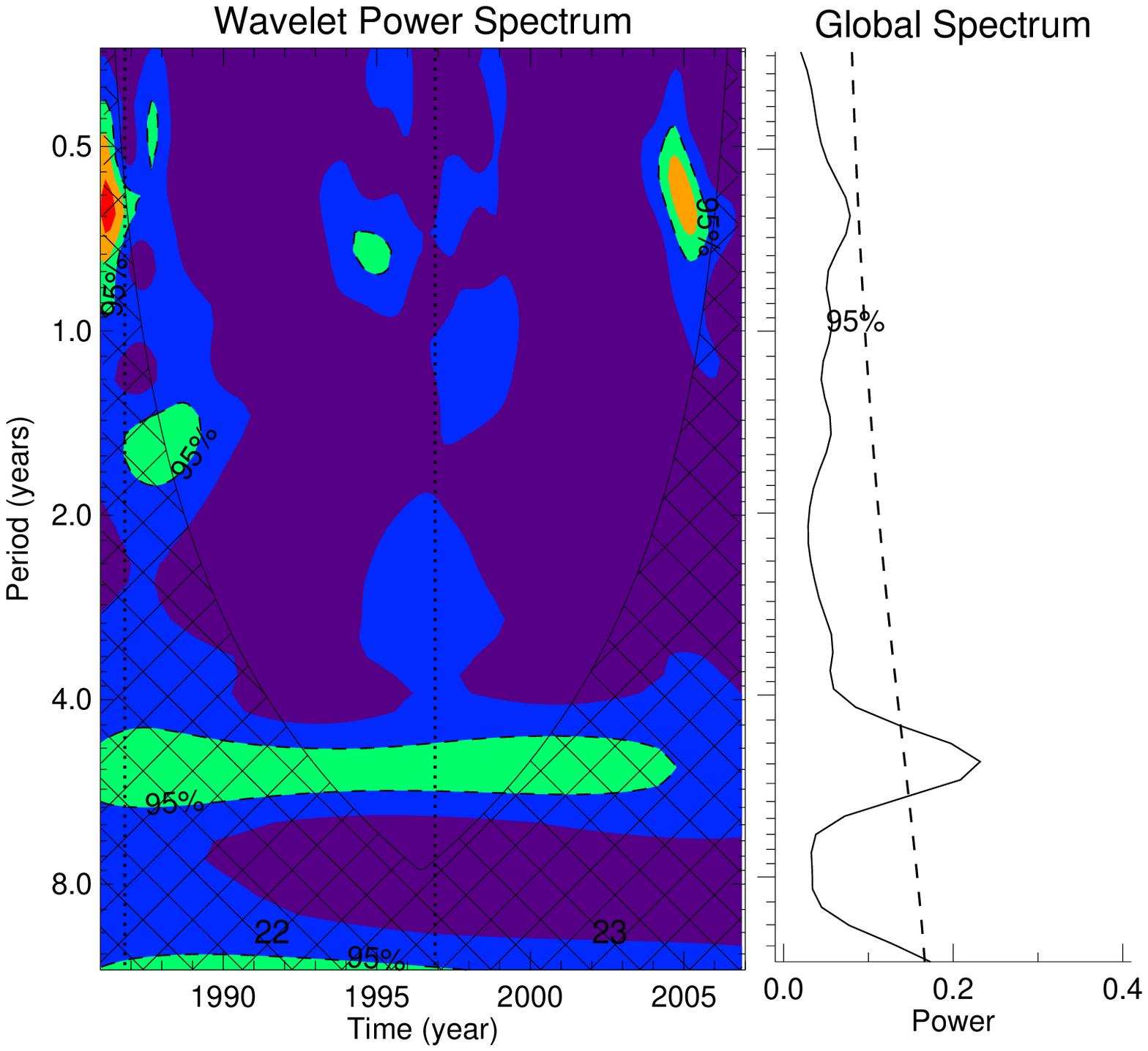}}
{\includegraphics[width=10.5cm]{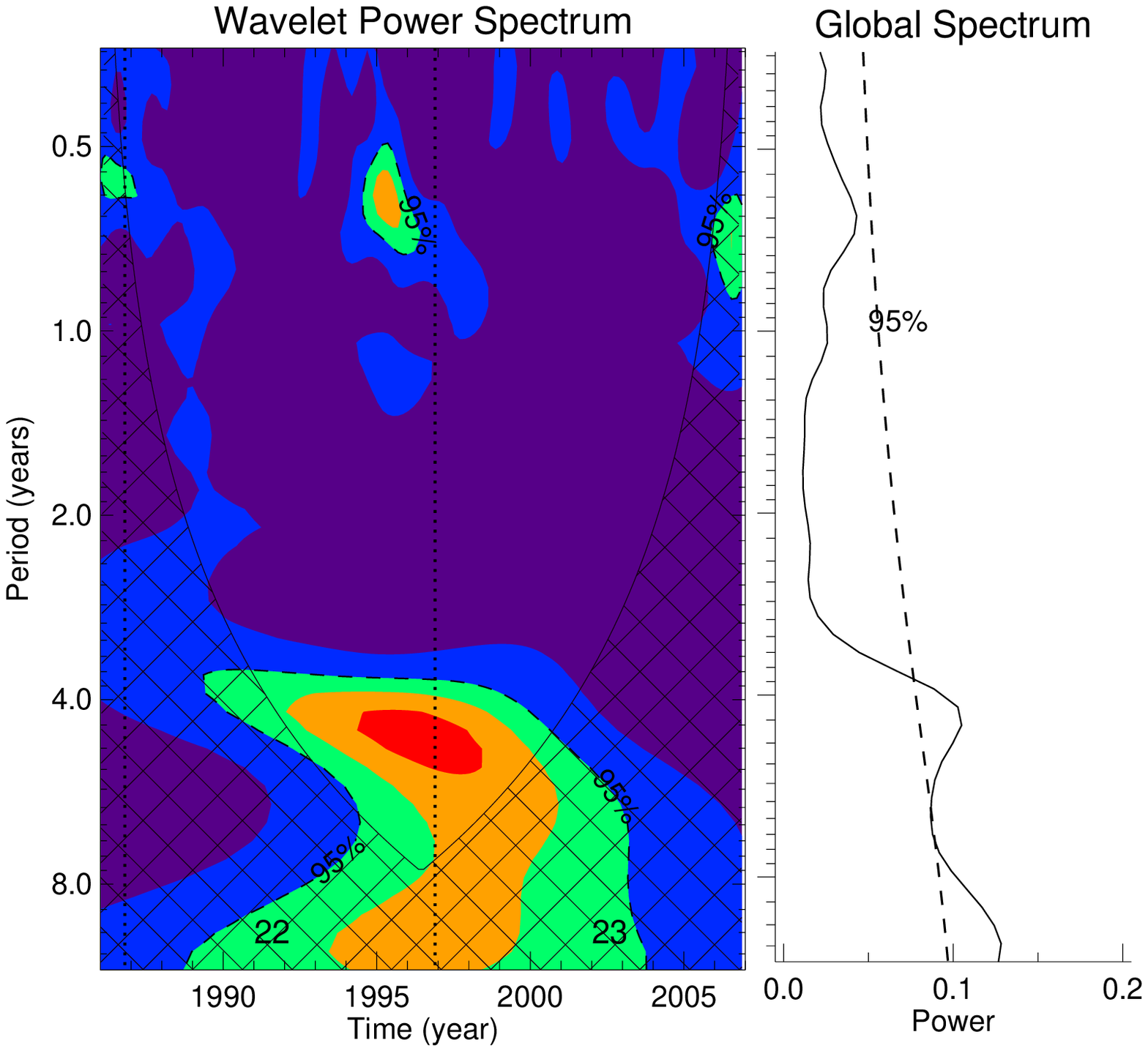}}
\caption{Wavelet power spectra and global spectra of $A$ determined from 
the time series of the corrected Mt. Wilson Doppler-velocity data (upper 
panel), SOON (middle panel) and DPD (lower panel) sunspot-group data.
 The wavelet spectra are  normalized 
by the variances of the corresponding time series. The shadings are  at
 the normalized variances of 1.0, 3.0, 4.5, and 6.0.
 The dashed curves represent the 95\% confidence levels
 deduced by assuming a white-noise process.
The cross-hatched regions indicate the cone of
influence where edge effects become significant (Torrance and Compo, 1998).
The dotted vertical lines indicate
the minima of the solar cycles. The Waldmeier number of the solar cycle    
is also given.} 
\label{Fig.4}
\end{figure}

\begin{figure}
\centerline{\includegraphics[width=\textwidth]{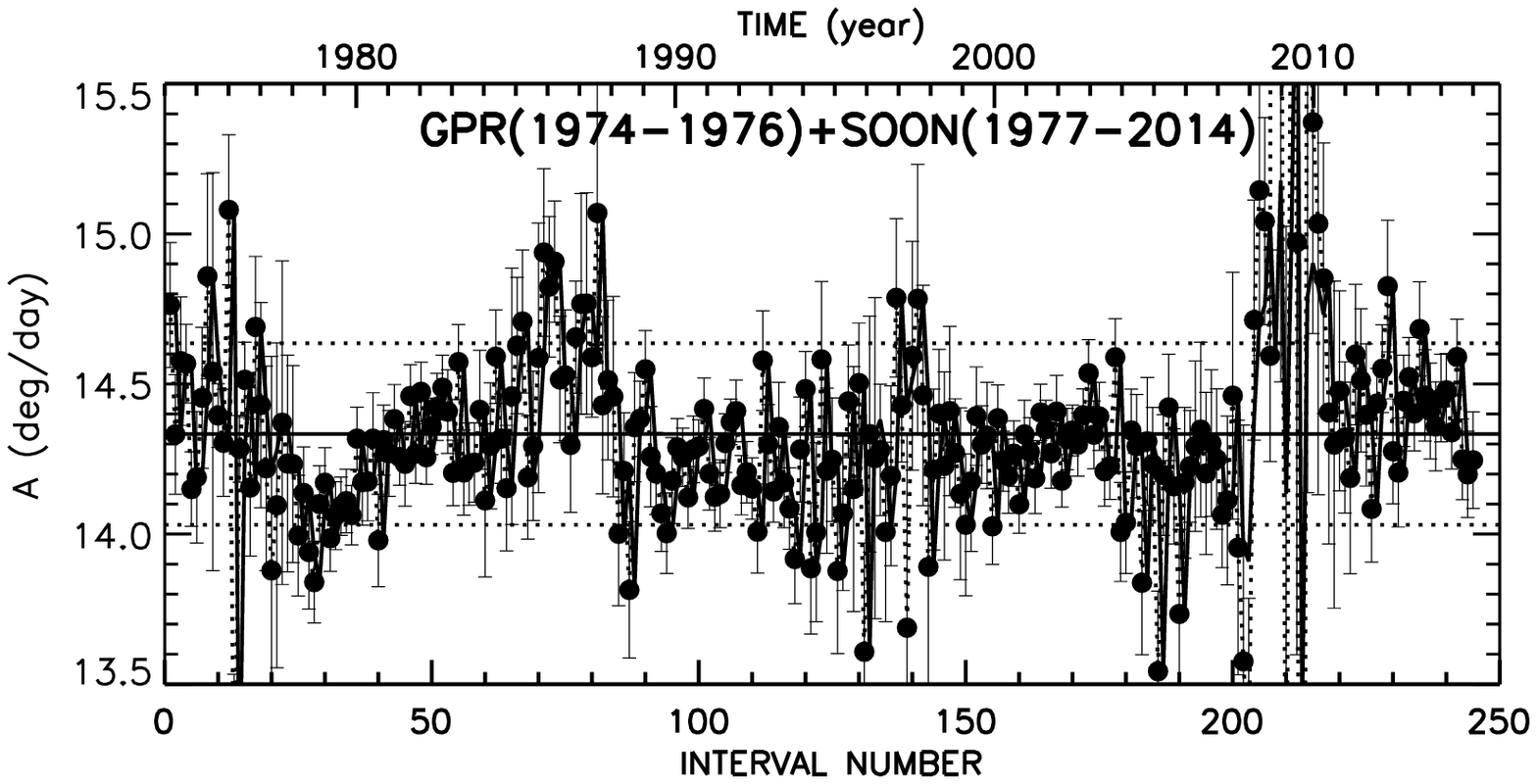}}
\centerline{\includegraphics[width=\textwidth]{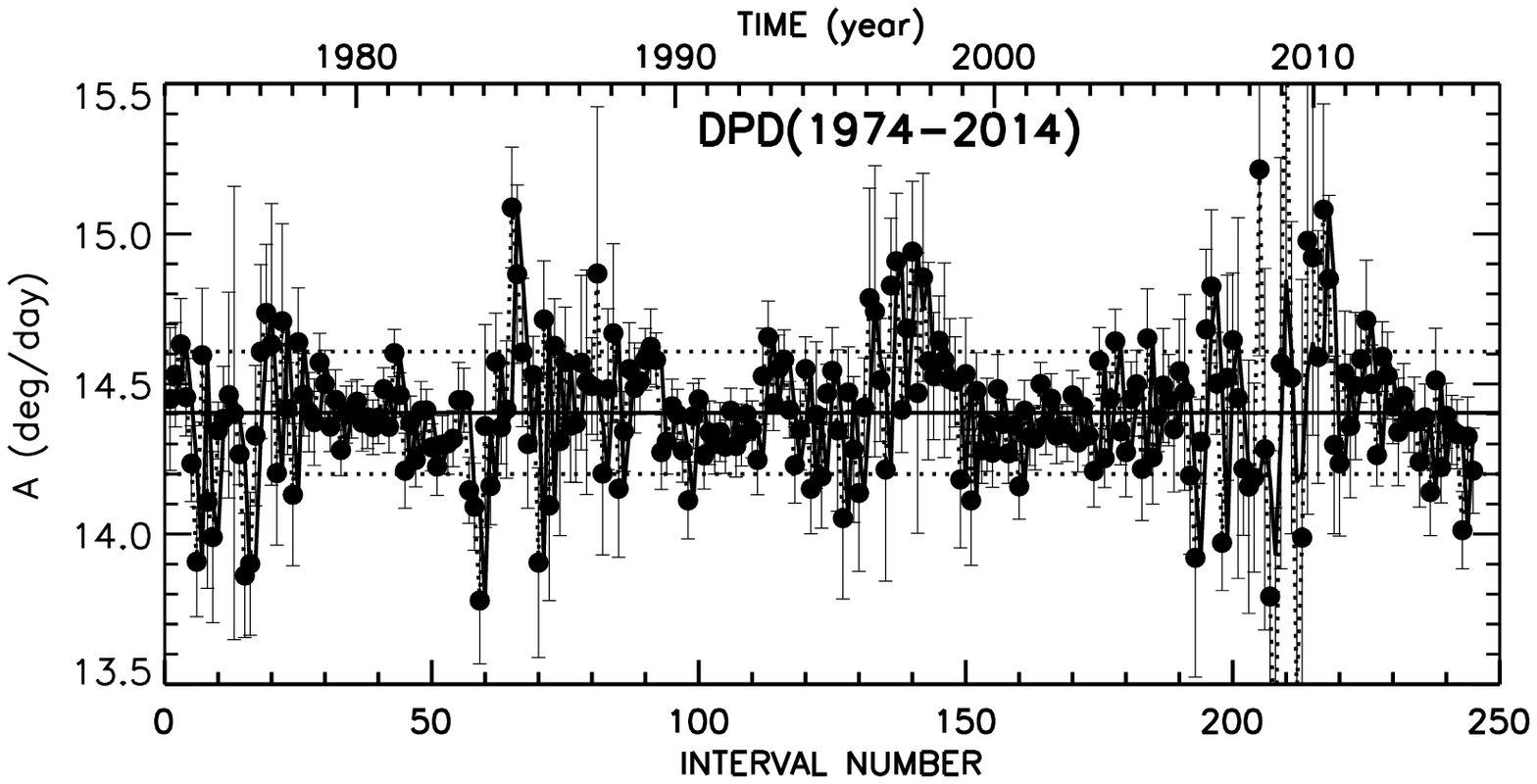}}
\caption{The lower and upper panels show plots of $A$ in 61-day
 intervals determined from
  the DPD (1974\,--\,2014) and the 
 combined GPR (1974\,--\,1976) and 
SOON (1977\,--\,2014) sunspot-group data
  $versus$ time (interval numbers), respectively. Error-bars represent
 the corresponding 1.0-$\sigma$ values.  
The horizontal continuous line represents the  mean and 
the horizontal dotted lines indicate the corresponding 
root-mean-square deviations.
The values whose
 $\sigma$ values  exceeded by 2.6 times the corresponding median value are 
replaced with the average of the corresponding values and their respective two
 neighbors.  The continuous curve represents the corrected data,  and 
the original data points are connected by the dotted curve.}
\label{Fig.5}
\end{figure}

\begin{figure}
\centerline{\includegraphics[width=\textwidth]{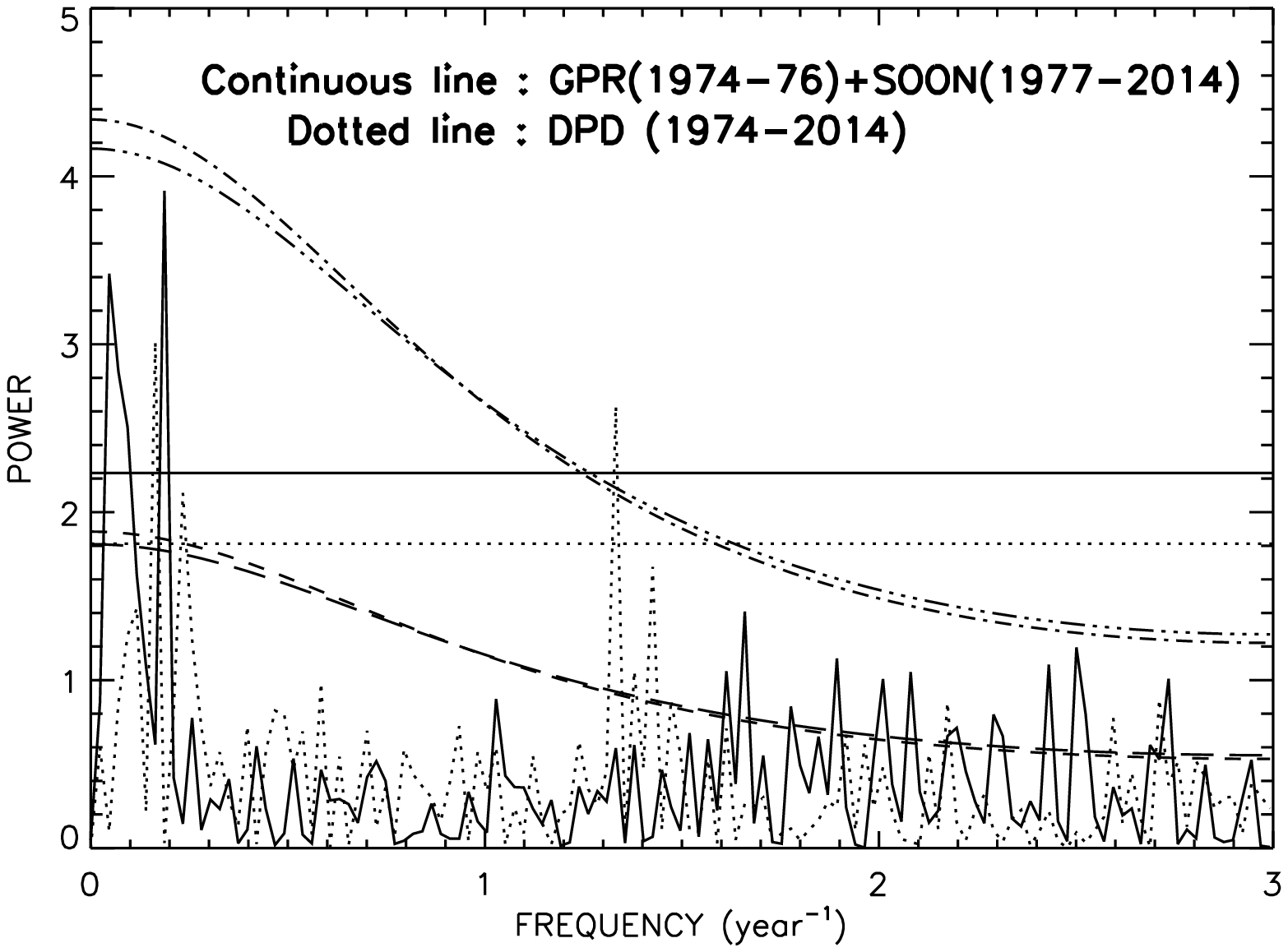}}
\centerline{\includegraphics[width=\textwidth]{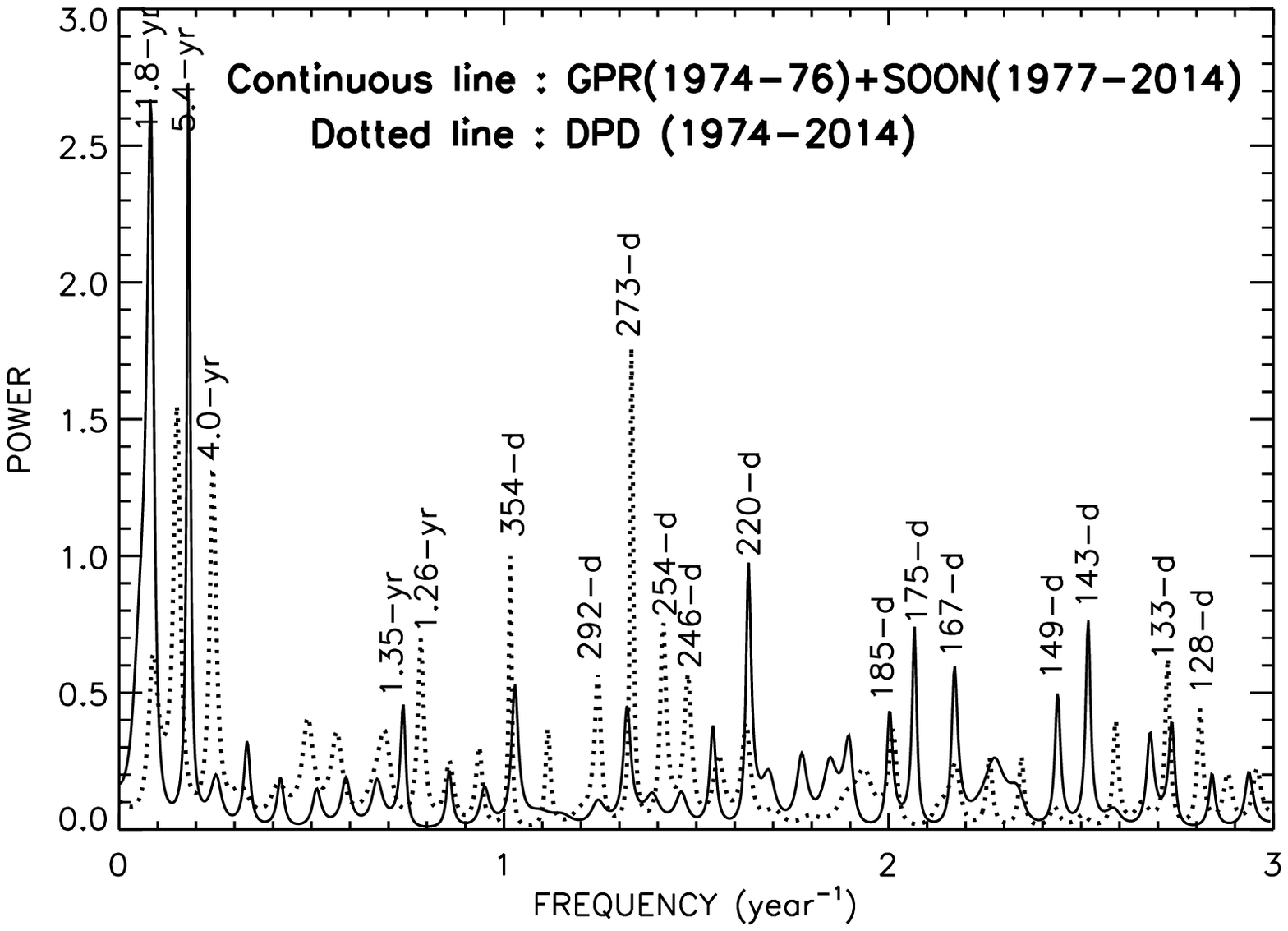}}
\caption{FFT  (upper panel) and MEM (lower panel)
 power spectra of $A$ determined from the  
corrected sunspot-group data shown in Figure~5.   
In the upper panel the long-dashed and  long dashed-dotted 
curves represent the mean and 90\% confidence level 
 red-noise spectra of $A$  determined from  the combined GPR and 
SOON  sunspot-group data during 1974\,--\,2014 ($\alpha = 0. 3065$), and the
dashed and  dash-dotted 
curves represent the corresponding spectra determined from DPD 
sunspot-group data 1974\,--\,2014 ($\alpha = 0.2870$).}
\label{Fig.6}
\end{figure}

\begin{figure}
\centerline{\includegraphics[width=\textwidth]{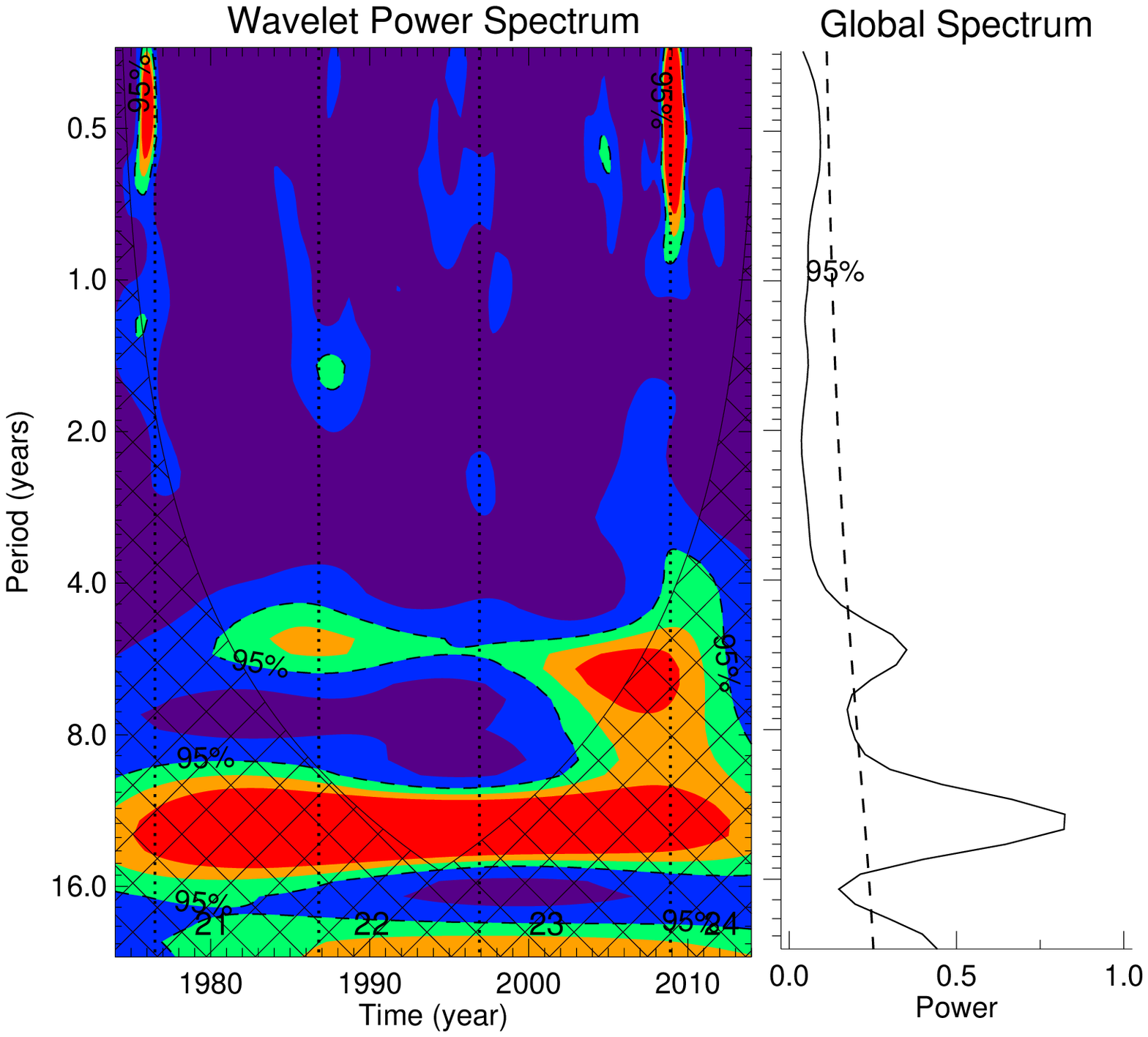}}
\centerline{\includegraphics[width=\textwidth]{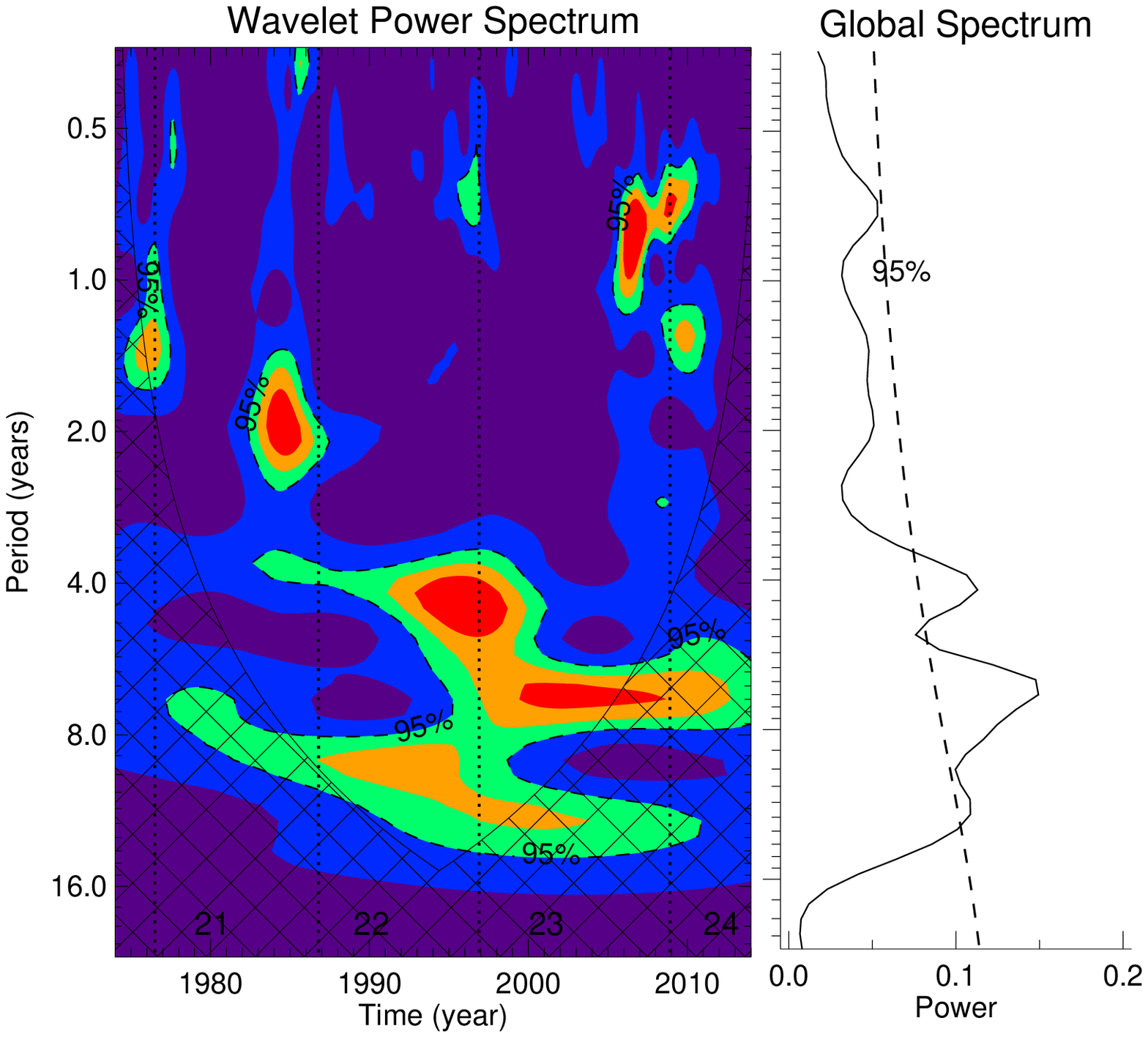}}
\caption{Wavelet power spectra and global spectra of $A$ determined from 
the combined GPR+SOON (upper panel) and
DPD (lower panel) sunspot-group data shown in Figure~5.   
 The wavelet spectra are  normalized 
by the variances of the corresponding time series. The shadings are  at
 the normalized variances of 1.0, 3.0, 4.5, and 6.0.
 The dashed curves represent the 95\% confidence levels
 deduced by assuming a white-noise process.
The cross-hatched regions indicate the cone of
influence where edge effects become significant (Torrance and Compo, 1998).
The dotted vertical lines indicate
the minima of the solar cycles. The Waldmeier number of the solar cycle    
is also given.} 
\label{Fig.7}
\end{figure}

\begin{figure}
\centerline{\includegraphics[width=\textwidth]{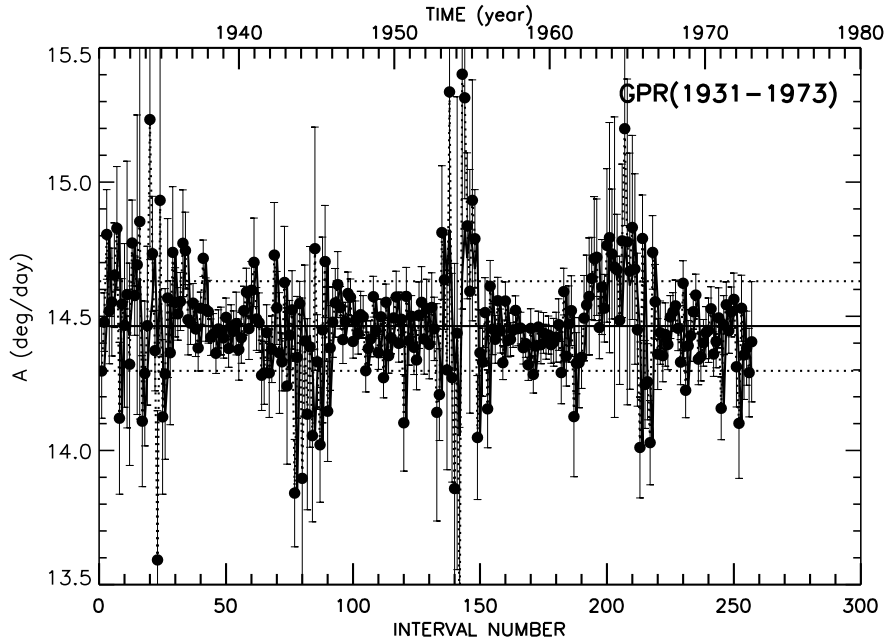}}
\caption{Plot of $A$ in 61-day intervals determined from
 the  GPR sunspot-group data during 1931\,--\,1973 $versus$ time
 (interval numbers).   Error bars represent
 the corresponding 1.0-$\sigma$ values.  
The horizontal continuous line represents the  mean and 
the horizontal dotted lines indicate the corresponding 
root-mean-square deviations.
The values whose
 $\sigma$ values  exceeded by 2.6 times the corresponding median value are 
replaced with the average of the corresponding values and their respective two
 neighbors.  The continuous curve represents the corrected data,   and 
the original data points are connected by the dotted curve.}
\label{Fig.8}
\end{figure}

\begin{figure}
\centerline{\includegraphics[width=\textwidth]{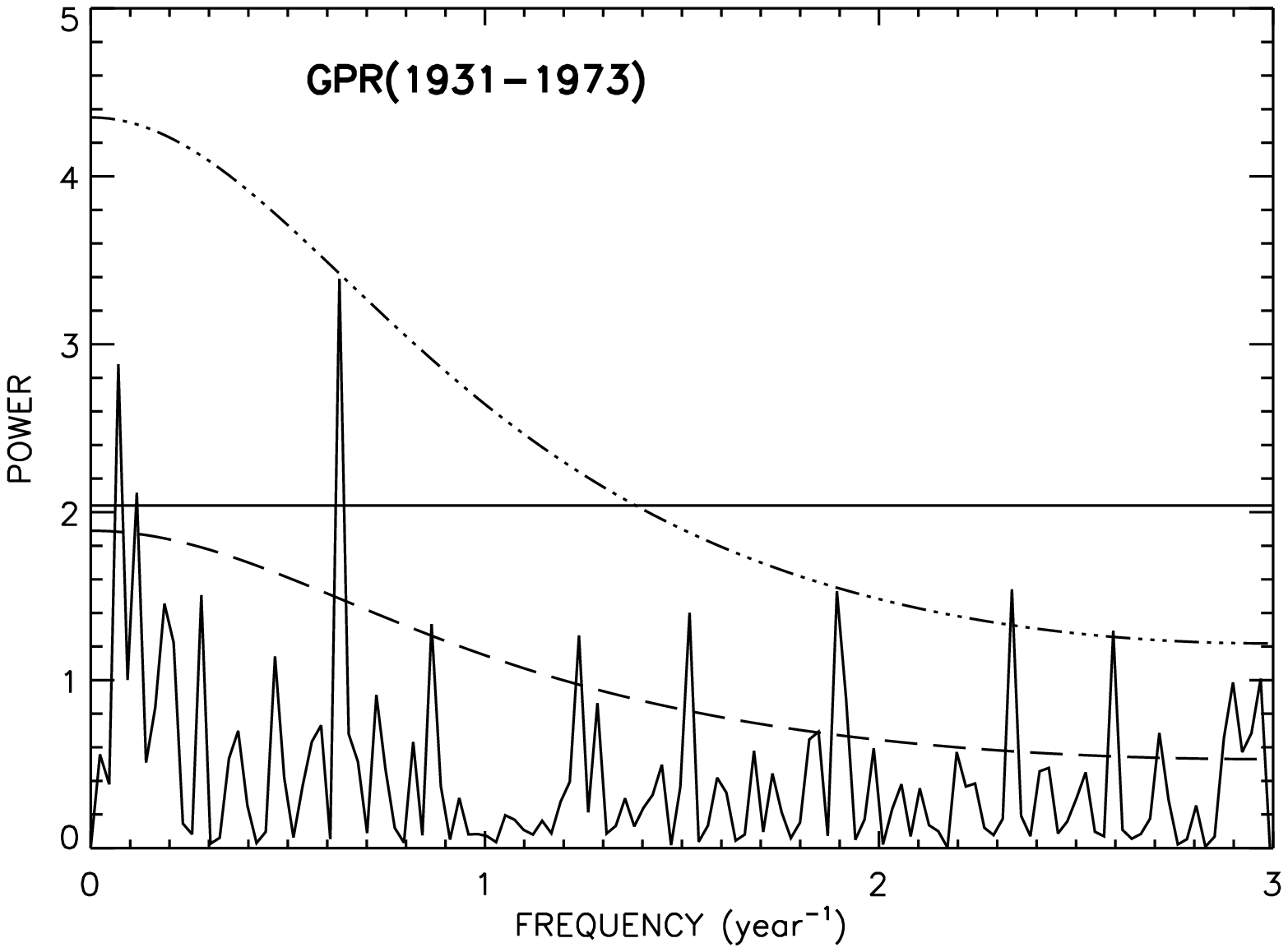}}
\centerline{\includegraphics[width=\textwidth]{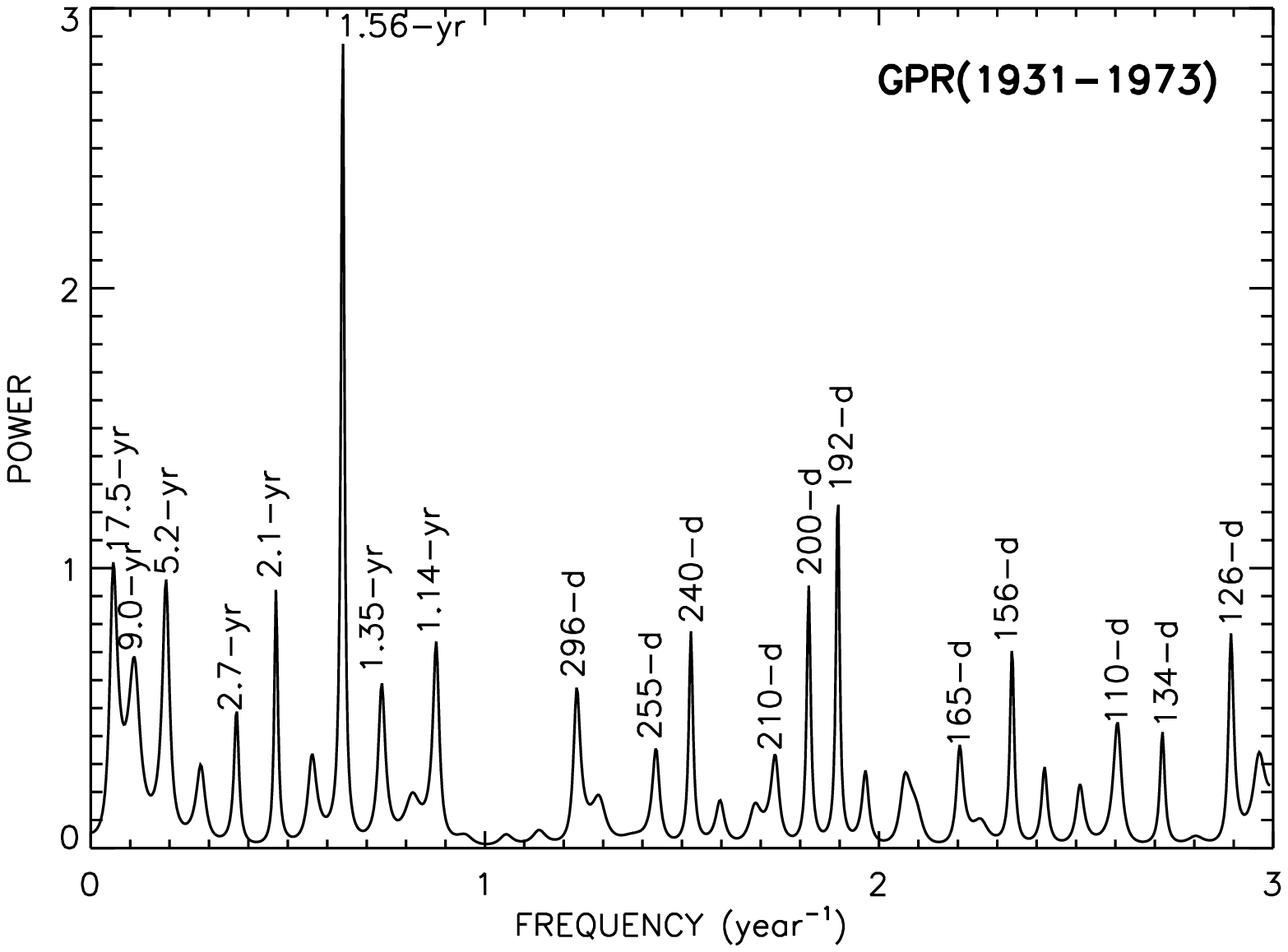}}
\caption{FFT  (upper panel) and MEM (lower panel)
 power spectra of $A$ determined from the  
corrected sunspot-group data shown in Figure~8.   
In the upper panel the long dashed and long dash-dotted 
curves represent the corresponding mean and 90\% confidence level 
 red-noise spectra ($\alpha = 0.308$).}
\label{Fig.9}
\end{figure}

\begin{figure}
\centerline{\includegraphics[width=\textwidth]{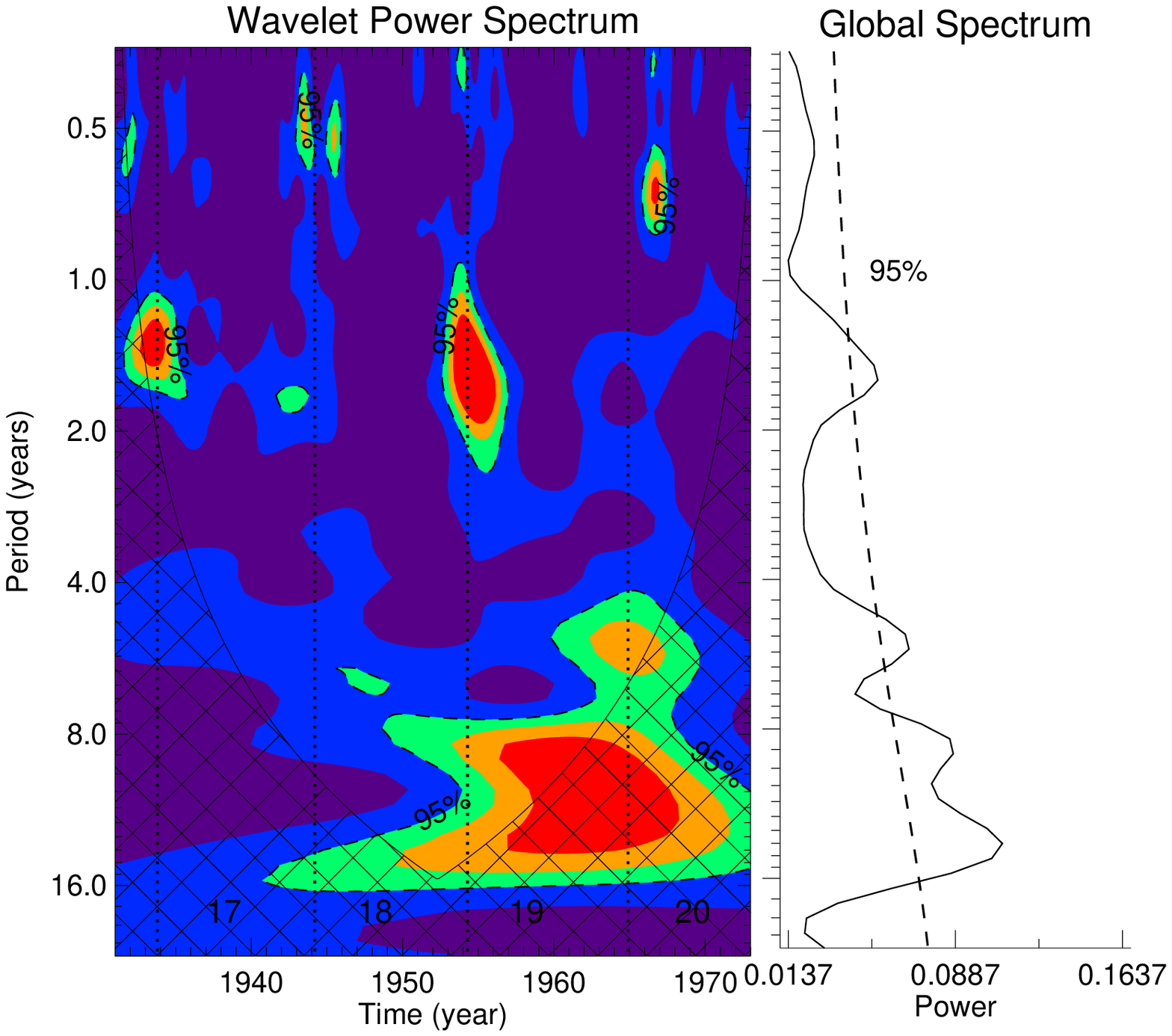}}
\caption{Wavelet power spectra and global spectra of $A$ determined from 
 the GPR  sunspot-group data shown in Figure~8.   
 The wavelet spectrum is  normalized 
by the variance of the corresponding time series. The shadings are  at
 the normalized variances of 1.0, 3.0, 4.5, and 6.0.
 The dashed curves represent the 95\% confidence levels
 deduced by assuming a white-noise process.
The cross-hatched region indicates the cone of
influence where edge effects become significant (Torrance and Compo, 1998).
The dotted vertical lines indicate
the minima of the solar cycles. The Waldmeier number of the solar cycle    
is also given.} 
\label{Fig.10}
\end{figure}

Figure~5 shows the variations in $A$ determined from the combined GPR 
and SOON sunspot-group data during the period 1974\,--\,2014 and DPD
 sunspot-group data during the same period. This figure shows that 
during 1974\,--\,1976 there is
a large difference in the values of $A$ determined from 
the combined GPR and SOON sunspot-group data  and the DPD
 data which is most likely due to the large uncertainties in $A$
during periods of minimum solar activity. In particular, during the last deep and prolonged 
minimum (between Cycles~23 and 24) the values are much more uncertain and the  
difference between the values of the SOON and DPD sunspot-group data
 seem to be even larger.  
 The correlation between the variations in  $A$ determined from 
the combined GPR and SOON  data and  DPD data is found to be only 14\%.  

Figure~6 shows the FFT and MEM power spectra, 
and Figure~7 shows the Morlet wavelet spectra of $A$ 
  determined from the combined GPR and SOON data  
 and DPD data shown in Figure~5.
Most of the peaks seen in the FFT and MEM spectra 
of $A$ determined from the sunspot group-data during 1986\,--\,2007 shown 
in Figures~2 and 3 are also present in the spectra shown in
 Figures~6 and 7.
 In the FFT spectrum of $A$ determined from the DPD sunspot-group
data during 1974\,--\,2014 the peak at frequency $~$1/273 day$^{-1}$ 
is significant 
at a 95\% confidence level in the white-noise model and at a 90\% confidence 
level in the red-noise model. There are peaks at $\sim$143 days and
 $\sim$175 days in the combined  GPR and SOON data, which are absent from 
  DPD data
  The peaks of $\sim$2.1-year and $\sim$1.44-year periodicities are
 poorly visible.     

In Figure~7, a strong signal with a 250\,--\,270-day periodicity exists 
during 2007\,--\,2010 in the $A$ data determined from the DPD sunspot-group
 data. Moreover, this periodicity seems to exist at regular intervals
 ($i.e.$  at the minimum of each solar cycle). 
This periodicity is to some extent also present in the wavelet
spectrum of  the combined GPR and SOON data. In addition, these data  
show the existence of $\sim$1.4-year periodicity 
around  1990, not identifiable 
in the DPD data. On the other hand,  there are strong 
signatures of this periodicity in the DPD data around 1976 and 2010,
 and a weaker one
close to 1995. Indication of such a periodicity can be also found  in
the combined GPR and SOON data during 1976. There is a very strong 
signal of a periodicity of $\sim$2 years  in the DPD data around 1985, 
while there 
is no signal of this periodicity  in the combined GPR and SOON 
data. There are weak signals (not well resolved) of this periodicity close
 to 1976 and close to 2008 in the combined GPR and SOON data. A $\sim$4.5-year 
periodicity exists during the period 1980\,--\,2012 in the combined GPR
 and SOON data, while  the DPD data show a periodicity $\sim$4-years
 during 1980\,--\,1998 and a strong periodicity of $\sim$5.4 years
  from 1998 onward. A 10\,--\,12-year periodicity seems too strong  in 
the combined GPR and SOON data throughout 1974\,--\,2014.
 A similar periodicity seems to exist in the DPD data, but the main 
 portion of the corresponding power is within  the cross-hatched regions 
where edge effects are significant. 
A similar behavior can also be seen in the wavelet spectrum of the DPD data.

Figure~8 shows the variations in the $A$ coefficient determined from the 61-day binned  GPR sunspot-group data during the period 
 1931\,--\,1973.  Figures~9 and 10 show
  the corresponding FFT, MEM, and Morlet wavelet power spectra.
Both the FFT and MEM spectra show similar sharp and well-defined peaks.
 The periodicity of $\sim$1.56 years is clearly visible in both of
 these spectra, above the 99\% confidence 
level using the  white-noise model and above the 90\% using 
the red-noise model. The FFT and MEM spectra also show a significant 
 $\sim$17.5-year  peak, which reaches the 99\% confidence level
 (white-noise model) in the FFT spectrum. 
This periodicity was also revealed in an earlier study of 
the sunspot group data by \cite{jj05}.  
As in Figures~2, 3, and 6,  many  other 
 peaks  are present in both spectra.
 The periods at  192 days, 156 days, 
 and 110 days periods are significant at a 90\% confidence level
 in the red-noise model.
It should be noted that,  as previously mentioned, most
 of the known short-term periodicities, $\le 5$ year,  in solar activity  
seem to appear only from time to time, $i.e.$ they are highly intermittent.  
The wavelet spectrum (Figure~10) shows a strong  signal
with a 1.56-year periodicity around  
1933 and 1955 that are also  visible  in the corresponding 
 FFT and MEM power spectra (Figure~9),  and  a strong 
$\sim$1.8-year periodicity around 1943. The periodicity at $\sim$296 days
  is clearly 
present near 1967, and the $\sim$200-day periodicity seems to 
appear around 1932, 1944, and 1946.  A weak signal with a 4\,--\,6-year 
periodicity throughout 1931\,--\,1973 becomes very strong around 1965.
The 10\,--\,17-year periodicity seems to be present only 
after 1940, particularly during the 1950\,--\,1970 time interval.

 Overall, the wavelet spectra of $A$ determined from the
 sunspot group data suggest that most of the short-term
 periods in $A$  appear around 
 the minimum of the solar cycles, where the uncertainties in the values
 of $A$ are
 substantial. The timing
 of the short-term periods in the velocity and sunspot-group data 
also disagrees considerably. 
The periods found  
  in the velocity data before 1995 may be artifacts due to frequent changes in 
the Mt. Wilson spectrograph instrumentation  (\opencite{jub09}). The
 abnormal longitudinal drifts ($>3^\circ$ day$^{-1}$)  of sunspot groups 
(\opencite{jj13}) were not excluded from our analysis. 
This may contribute to the uncertainties in the periods found in $A$ from
 the sunspot-group data.  In addition, the evolution of sunspot groups may
 also play a significant role. 
 These problems raise some questions about the existence and/or
 significance of the short-term periodicities
found here in the $A$ time series. However, these periodicities have
also  been found in other solar activity indices. For example, \inlinecite{gurg16}  have 
detected similar signatures in
 selected datasets during a few years of the solar cycles and have shown that
 their length and level of significance
vary with time. This suggests that the short-term periodicities found in our 
investigation are worth  further
study to prove their existence.

\section{Conclusions and discussion}
Our analysis indicates that short-term variations in the
solar equatorial rotation rate are highly intermittent in nature.
A period of $\sim$250 days  is found in the variations of the equatorial
rotation rate determined from both the
Mt. Wilson Doppler-velocity data and the sunspot-group data.
In the equatorial rotation rate of sunspot groups  a strong $\sim$1.4-year
period is found  around 1956, and weak signals of this period
are seen also around 1935, 1976, 1986, and 1996.  The
 velocity data  show a weak signature of this periodicity 
during 1990\,--\,1995 in these measurements.
A  strong periodicity of $\sim$1.6 years  is found
during  1933 and 1955 in the equatorial
rotation rate of sunspot groups. A weaker signal of the same period
is seen  around 1976, while a strong $\sim$1.8-year
 periodicity is found around 1943.
There are indications that periods of $\sim$5.4 years, $\sim$11 years,
and $\sim$17.5 years  exist in the equatorial rotation
 rate of sunspot groups, while a $\sim$7.6-year period
 is found only in the velocity data.  In the sunspot data there
 is also a suggestion of  
 several  other short-term 
periods, $viz.$, $\sim$182 days (around 1987, 1995, and 2005), 
$\sim$200 days (around 1932, 1944, and 1946), $\sim$1 year, 
$\sim$2 years, $\sim$4 years, $etc.$  
However, the actual existence of most of the short-term  periodicities 
found here needs to be confirmed  because they might be artifacts of the 
large uncertainties in the data that are due to various factors, 
for example, frequent changes in
the Mt. Wilson spectrograph instrumentation,  
abnormal longitudinal drifts ($>3^\circ$ day$^{-1}$)  of sunspot group, 
evolution of sunspot groups.

Since on average the rotation rate of sunspot groups  (in  general
 magnetic regions)  represents the 
rotation rate of somewhat deeper layers of the Sun  depending upon the sizes 
of the groups (\opencite{fouk72}; 
\opencite{gf79}; \opencite{nesme93}; \opencite{how96}; \opencite{jk99};
 \opencite{jg97b};  \opencite{hi02}; \opencite{siva03}; \opencite{jj13}), 
the $\sim$1.4-year periodicity in the  equatorial rotation rate of
 sunspot groups  may be   related to the known 1.3-year periodicity 
in the low-latitude rotation rate at the base of the convection zone 
(\opencite{howe00}).  
The physical connection between a periodicity in the solar rotation and the 
corresponding periodicity in solar activity may be explained as follows:  
 the variation in the emerging magnetic flux on the Sun could be
 affected by the corresponding modulation in the solar rotation, through  
 the  effect of the Coriolis force on the emerging
 magnetic flux. However, the causes of the variation in 
the solar rotation are unknown. 
One possible explanation is that the variation in solar differential rotation 
may be caused by  Rossby-type waves as discussed by 
\inlinecite{war65}, \inlinecite{kuhn98}, and \inlinecite{ksb05}.
Another possibility is that short-term periodicities may be caused by internal
 gravity waves as discussed by \inlinecite{wolff83}, or 
 that the source of the perturbation may come from outside the Sun, $i.e.$
 from solar system dynamics (\opencite{wood65}; \opencite{jg95}; 
 \opencite{zaq97}; \opencite{juck00}; \opencite{jj03}, 2005;  
\opencite{wil08}; \opencite{mhg10}; \opencite{wolff10}; \opencite{tan11};
 \opencite{cc12}; \opencite{abreu12}; \opencite{wil13};
 \opencite{chow16} and references therein). In the latter case, 
it is interesting to note
 that the periodicities of $\sim$250 days and  $\sim$1.6 years are  
 approximately the orbital period of the planet Venus and its
 modulation  caused by relative position of Earth and Jupiter.
  All these are important planets because their tidal forces are
 relatively strong.   Moreover, the combined
 effect of these planets  seems to be strongest at the times that are at or
 near solar cycle minima~(\opencite{wil13}). This might be a 
 reason for  the  short-term periodicities found  around solar cycle minima.
  Nevertheless,  we cannot rule out the possibility that  the
 variations in the rotation rate of sunspot groups simply represent    
  the variations in the sunspot emergence along with the 
group size and longitudinal displacement.

 The periodicities of 182 days and one year periodicities are mostly
 artifacts of the seasonal
 effects, but they may be also   
related to the relative  positions of  Earth and Jupiter.
 Most of the short-term periodicities may be related to 
relative positions of  Mercury, Venus, Earth, Mars and 
Jupiter. However,  mere matching of  
periodicities is no guaranty for a physical relationship. 
The  periodicity of $\sim$2 years may be related to the well-known
quasi-biennial oscillation of the solar activity, and it may be related to 
the configuration of Earth, Mars, and Jupiter (the Earth-Mars
synodic period is 2.136 years).
Any pair of planets will be 
in alignment and produce relatively high suntides periodically at 
intervals equal to one-half their mean synodic period. Therefore, 
the origin of other relatively long-term periodicities 
can also be explained on the basis of alignments of  Venus and Earth, 
relative to the positions of Jupiter (also see \opencite{wil13}). 
That is, the periodicities of $\sim$5.4 years, 
$\sim$7.6 years, and $\sim$17.5 years found here may be 3.5,
 4.5, and  11 times the synodic period (1.597 year) of  Venus and
 Earth. In an earlier analysis the average  solar cycle 
variation of the  equatorial rotation rate was found to be strong  
  during an odd-numbered solar cycle and the variation was  weak or absent 
during an even-numbered solar cycle (\opencite{jj03}). 
The $\sim$17.5-year period may be related to this property. 
  The peak  of the $\sim$7.6-year period in the power spectra of  the
 velocity data may be an artifact of the large difference in 
  these data 
 before and after 1995. However, a 7-year periodicity is known to exist 
in the solar rotation rate derived from Ca II K plage data during the period
 1951\,--\,1981 (\opencite{jsb85}). The origin of all of the detected 
 periodicities still needs to be established.   

\acknowledgements{We thank the anonymous referee for the critical review and 
useful comments and suggestions. Wavelet software was provided
 by C. Torrence and G. Compo
 and is available at {\tt http//paos.colorado.edu/research/wavelets}. 
The MEM software was provided by Dr. A. V. Raveendran.}
 
{\bf Disclosure of Potential Conflict of Interest} The authors declare
 that they have no conflicts of interest.

{} 
\end{article}
\end{document}